\documentclass[fleqn,usenatbib]{mnras}

\usepackage{newtxtext,newtxmath}

\usepackage[T1]{fontenc}

\DeclareRobustCommand{\VAN}[3]{#2}
\let\VANthebibliography\thebibliography
\def\thebibliography{\DeclareRobustCommand{\VAN}[3]{##3}\VANthebibliography}

\usepackage{graphicx}	
\usepackage{amsmath}	
\usepackage{physics}    
\usepackage{multirow}

\newcommand{\invday}{\,d$^{-1}$}
\newcommand{\h}{^\text{h}}
\newcommand{\m}{^\text{m}}

\newcommand{\electronpersecond}{e$^-\,$s$^{-1}$}

\title[Tilted discs in six poorly studied CVs]{
    Tilted discs in six poorly studied cataclysmic variables
}

\author[Stefanov \& Stefanov]{
Stefan Y. Stefanov,$^{1,2}$\thanks{E-mail: sstefanov@nao-rozhen.org}
Atanas K. Stefanov$^{3}$
\\
$^{1}$Institute of Astronomy and National Astronomical Observatory,
Bulgarian Academy of Sciences, 72 Tsarigradsko Shose Boulevard, 1784
Sofia, Bulgaria \\
$^{2}$Department of Astronomy, Sofia University "St. Kliment Ohridski",
5 James Bourchier Boulevard, 1164 Sofia, Bulgaria \\
$^{3}$Department of Physics and Astronomy, University College London,
Gower Street, London WC1E 6BT, UK
}

\date{Accepted XXX. Received YYY; in original form ZZZ}

\pubyear{2022}

\begin{document}
\label{firstpage}
\pagerange{\pageref{firstpage}--\pageref{lastpage}}
\maketitle

\begin{abstract}

In this work, we search for negative superhumps (nSHs) in poorly studied
cataclysmic variables using TESS data. We find three eclipsing binaries
with nSH signatures: HBHA 4204-09, Gaia DR3 5931071148325476992, and
SDSS J090113.51+144704.6. The last one exhibits IW~And-like behaviour in
archival ZTF data, and appears to have shallow, grazing eclipses. In
addition, we detect nSH signatures in two non-eclipsing systems: KQ Mon
and Gaia DR3 4684361817175293440, by identifying the orbital period from
the superorbital-dependent irradiation of the secondary. We discover nSH
signatures in one more system, [PK2008]~HalphaJ103959, by using an
orbital period from another work. An improved mass ratio -- nSH deficit
relation $q(\varepsilon_-)$ is suggested by us, which agrees with
independent measurements on nova-like variables. With this relation, we
estimate the mass ratios of all systems in our sample, and determine the
orbital inclinations for the three that are eclipsing. All systems with
discovered nSHs in this work are excellent targets for follow-up
spectroscopic studies.

\end{abstract}




\section{Introduction}

Cataclysmic variables (CVs) are binary systems that consist of a
white-dwarf (WD) primary and a Roche-lobe filling secondary. Matter from
the secondary flows through the first Lagrangian point and accretes on
to the primary. In the case of a non-magnetic or a very weakly magnetic
primary, this mass transfer happens through an accretion disc
\citep{Hellier2001}. In systems with mass-transfer rates of
\mbox{$\dot{\text{M}} \simeq$ 1 -- 5 $\times 10^{-9}$
M$_\odot$yr$^{-1}$}, thermal instabilities arise in the accretion disc
and cause repeating quasi-periodic outbursts. These outbursts usually
occur once about every few months, last several days, and can increase
the system brightness with up to \mbox{$\sim$~5 mag}. CVs with recorded
outbursts are termed dwarf novae (DNe), whereas CVs with no recorded
outbursts are termed nova-likes (NLs). In NLs, most of the flux
originates from the accretion disc, which is in a hot steady state and
is much brighter than the two system components. The orbital periods of
this type of variables can range from $\sim 1$~h to more than 10~h. Not
many CVs, however, are observed in the period range of 2--3~h. This
phenomenon is called the ''period gap'' and is explained by transitions
in evolutionary stages of this type of variables (see
\citealp{Warner1995} for an encyclopedic description of CVs).

NLs can change their brightness on time-scales from seconds to
millennia. Some systems have drops in brightness of several magnitudes,
which can last from months to years. This behaviour is most commonly
observed in systems with orbital periods $(P_\text{orb})$ near the upper
edge of the period gap. Such drops in brightness are categorised as a
low state of type VY~Scl \citep{King1998} and can also be displayed by
magnetic CVs. VY~Scl low states are likely caused by the reduction or
the complete cessation of mass transfer in the system, which
significantly decreases the flux coming from the disc. They are believed
to be associated with the magnetic activity of the secondary. Star spots
emerging on the first Lagrangian point may suppress mass transfer in the
system \citep{LivioPringle1994}, and the radius of the secondary itself
can be affected by magnetic activity \citep{Howell2004}. Yet, the exact
mechanism of mass-transfer cessation during VY~Scl episodes remains
unknown.  

Apart from low states and other long-term trends in brightness, CVs
display an abundance of photometric variability on shorter time-scales.
Roche-lobe geometry requires that the secondary takes a characteristic
teardrop-like shape. As it orbits the barycentre, it presents different
projections of itself to the observer, which introduces a photometric
variability of period $P_\text{orb}/2$. A similar effect can occur when
the secondary is strongly irradiated by the accretion disc. In that
case, the visibility of the irradiated side of the secondary is
dependent on the orbital phase of the system, and a light-curve
modulation of period $P_\text{orb}$ takes place.

Some CVs exhibit variations in brightness that have periods slightly
shorter or slightly longer than $P_\text{orb}$. These variations are
called ''superhumps'' and are believed to be caused by a precessing
accretion disc. Superhumps can be of either positive (pSH) or negative
(nSH) type, depending on the sign of \mbox{$P_\text{SH} -
P_\text{orb}$}. They are well-studied and commonly seen in SU~UMa stars,
a DN subclass (e.g. \citealp{Kato2009,Kato2017}); as well as in NLs
\citep{Bruch2023}. For NLs in particular, \citeauthor{Bruch2023} gave a
sample of 46 systems, 13 have had pSHs, 16 have had nSHs and 17 have had
superhumps of both types at some point in the past (but not necessarily
at the same time).

Each superhump type is associated with processes of different nature.
pSHs are believed to be caused by an apsidally precessing accretion
disc. In this case, the 3:1 resonance induces tidal deformations, the
heat from which causing periodic changes in disc brightness
\citep{Whitehurst1988,HiroseOsaki1990,Lubow1991}. On the other hand,
nSHs can be explained with a retrograde nodal precession of a tilted
accretion disc. The tilt allows for the mass-transfer stream to go
deeper in the gravitational well of the primary, and thus to release
more energy upon impact. The point of impact on the disc is commonly
referred to as the ''bright spot''. The sweeping of the bright spot
across the disc faces introduces an additional photometric variability
that has a period equal to the beating of $P_\text{orb}$ and the disc
precession period $P_\text{prec}$ \citep{Wood2009, Montgomery2009}, i.e.
\begin{equation}\label{eq:beat}
    \dfrac{1}{P_\text{nSH}} =
    \dfrac{1}{P_\text{orb}} + \dfrac{1}{P_\text{prec}}.
\end{equation}

Superhumps of both types can be used to estimate some physical
properties of these systems. The nSH deficit $\varepsilon_-$ is defined
as
\begin{equation}
\label{eq:epsilon-}
\varepsilon_- = \dfrac{P_\text{nSH} - P_\text{orb}}{P_\text{orb}}
\end{equation}
and has been shown to correlate with the mass ratio of the system
\mbox{$q=M_1/M_2$} in several works (e.g. \citealp{Wood2009,
Montgomery2009}). A detailed study of nSHs can be found in
\citet{Kimura2020,Kimura2021}, where Kepler photometry of the NL system
KIC~9406652 was analysed. The light curve of this particular object has
identifiable $P_\text{orb}$, $P_\text{nSH}$ signals as well as
superorbital ones (i.e. $P_\text{prec}$ signatures).

In this work, we present our results from a search for nSHs in poorly
studied CVs that are similar to KIC~9406652. Section~\ref{sec:Analysis}
presents our methods for searching and data reduction, and gives a list
of objects with discovered nSH signatures. Section~\ref{sec:Results}
contains a literature review and discussion of each system we found to
have nSH behaviour. In Section~\ref{sec:Discussion}, we attempt to
estimate some physical parameters in said systems, and in
Section~\ref{sec:Conclusions}, we summarise the findings of this work.

\section{Analysis}\label{sec:Analysis}

\subsection{Data from TESS}

The Transiting Exoplanet Survey Satellite (TESS; \citealp{tess}) mission
is an all-sky survey in the red-infrared that continues to provide with
long-term measurements of remarkable photometric precision. The TESS
Science Processing Operations Center pipeline (SPOC; \citealt{SPOC})
offers light curves from two different reduction techniques: Simple
Aperture Photometry (SAP) and Pre-Search Data Conditioning Simple
Aperture Photometry \mbox{(PDCSAP)}. A comprehensive comparison between
the two is given in \citet{Kinemuchi2012}. PDCSAP tries to reduce
effects of instrumental nature, but can sometimes introduce systematics
in periodograms, and analysis should proceed with care.
\citet{Bruch2022} found in particular that the additional conditioning
in PDCSAP may distort DNe light curves, and chose to use the simpler SAP
technique in order to search for periodic variations in CVs. We use SAP
light curves too in all analysis to follow.

\subsection{Photometric features of tilted accretion discs}\label{sec:superhump_degeneracy}

Negative superhumps are direct evidence for a titled accretion disc, but
finding their signatures is only possible in systems of known
$P_\text{orb}$. This is a strong restriction, since not many CVs have
had their orbital periods measured. To expand the population of stars
with known $P_\text{orb}$, we searched for systems with several
significant peaks in the power spectrum, in a frequency region above the
period gap. In the case of two neighbouring prominent peaks, it could be
that those are signatures of \mbox{$P_\text{pSH},P_\text{orb}$} and not
\mbox{$P_\text{nSH},P_\text{orb}$}. Nevertheless, this degeneracy can be
lifted with the following rationale.

\begin{figure}
    \centering
    \includegraphics[width=\columnwidth]{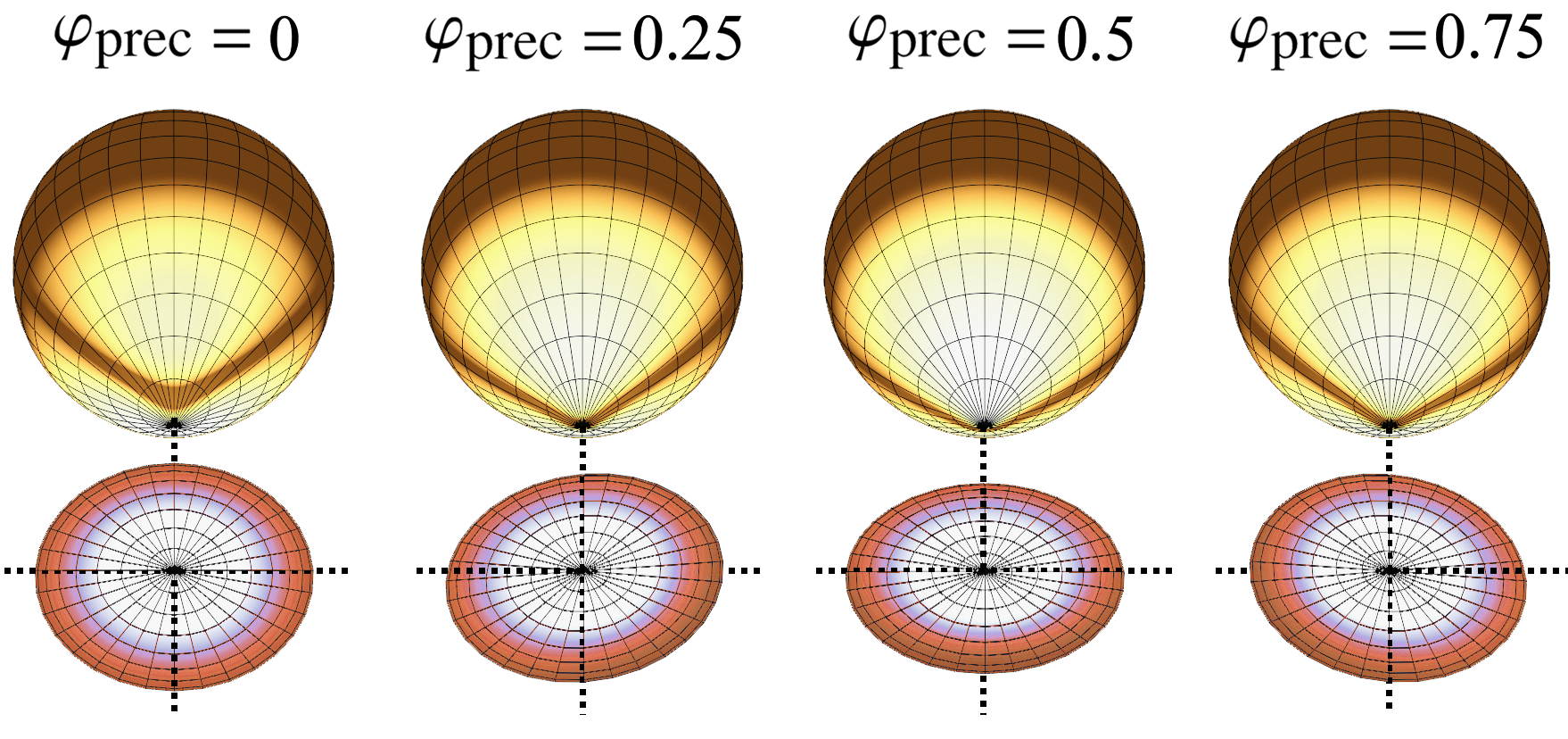}
    \caption{A model CV system at an orbital phase
    $\varphi_\text{orb}=0.5$, as it would be seen by an observer. Four
    precession phases $\varphi_\text{prec}$ of a disc with tilt
    $\theta=6\degr$ are illustrated. The orbital plane of the system is
    defined by dotted lines. It divides space into two half-spaces: the
    near (above the plane) and the far one (below the plane), with
    respect to the observer. The precession phase is defined such that
    the system is the brightest at \mbox{$\varphi_\text{prec} = 0$}. In
    this orientation, the disc has the largest projected area at
    \mbox{$\varphi_\text{prec}=0$}. Conversely, at
    \mbox{$\varphi_\text{prec} = 0.5$}, it has the smallest projection,
    but faces towards the secondary and irradiates it the most.
    \citet{Kimura2021}, Figure 9 gives a full description of CV
    configurations in titled-disc regimes.} 
 \label{fig:irradiation}
\end{figure}

In systems with a precessing tilted accretion disc, the disc orientation
changes with respect to the secondary for different orbital phases
$\varphi_\text{orb}$ and different disc precession phases
$\varphi_\text{prec}$. The former is defined such that the secondary is
at inferior conjunction at $\varphi_\text{orb}=0.0$; the latter is
defined\footnote{These definitions are consistent with
\citet{Kimura2020}.} such that the light maximum in the disc precession
cycle is at $\varphi_\text{prec}=0.0$. The observed irradiation of the
secondary by the bright disc varies with both $\varphi_\text{prec}$ and
$\varphi_\text{orb}$. Consider a non-eclipsing system at
$\varphi_\text{orb}=0.5$ (Figure~\ref{fig:irradiation}). The orbital
plane of the system divides space into two half-spaces, one of which the
observer finds themselves in. One part of the system resides in the same
half-space as the observer, and the other part is in the opposite
half-space (i.e. on the other side of the orbital plane with respect to
the observer). We shall refer to those as the ''near half-space'' and
the ''far half-space''. As an example, in Figure~\ref{fig:irradiation},
the part of the accretion disc that lies in the near half-space is: its
rear side at $\varphi_\text{prec}=0$, its right side at
$\varphi_\text{prec}=0.25$ and so on.

For an observer, the near half-space of a system is more photometrically
accessible than the far half-space.\footnote{This is only untrue in the
special case of $i=90\degr$, when the observer lies in the orbital
plane, and both half-spaces are thus equally accessible.} At
$\varphi_\text{prec}=0$, the luminous disc reveals the most of itself to
the observer, and the average system brightness across
$\varphi_\text{orb}$ is the greatest. However, the irradiated region of
the secondary is in the far half-space, and thus the
$\varphi_\text{orb}$ variation in brightness is minimal in amplitude. In
the opposite case of $\varphi_\text{prec}=0.5$, the observer sees the
smallest possible projection of the disc, and the average system
brightness across $\varphi_\text{orb}$ is the smallest -- but the
irradiated region of the secondary is now in the near half-space, and
the $\varphi_\text{orb}$ variation in brightness is maximal in
amplitude.

At the same time, nSHs introduce additional complexities in variability
that need to be accounted for. \citet{Kimura2021} discuss this issue and
carry out the following procedure. A given light curve is initially
split into subsets of different time intervals. Then, for each subset,
they: (1) fold by $P_\text{nSH}$ and construct an average light-curve
profile of the nSH, (2) subtract said profile from each subset, (3)
split the subset into different $\varphi_\text{prec}$ windows, (4) fold
each window by $P_\text{orb}$. This technique results in multiple
orbital phase curves, each corresponding to a different
$\varphi_\text{prec}$ window. If these phase curves show a
\mbox{$\varphi_\text{prec}$-dependent} irradiation of the secondary, the
system has a precessing tilted accretion disc and the observed superhump
is negative. It is this consideration that could lift the pSH-nSH
degeneracy in the power spectrum.

In order to address the nSH contamination, we use a variant of the
nSH-subtraction technique by \citet{Kimura2021} with the following
adjustments: all data is smoothed by a fourth-order Savitzky-Golay
filter \citep{Savitzky1964} of window size 10 d, and no separate subsets
are considered; in (1), nSH light-curve profiles are constructed with
median filters of window size $1101$;\footnote{We found that this window
size worked generally well for all systems.} in (3), four
$\varphi_\text{prec}$ intervals are considered with centres at
\mbox{$\varphi_\text{prec}=0.00,0.25,0.50,0.75$} and of width
0.1.\footnote{That is, the intervals \mbox{0.95 -- 0.05}, \mbox{0.20 --
0.30}, \mbox{0.45 -- 0.55},  \mbox{0.70 -- 0.80}.}

\subsection{Target selection}

The International Variable Star Index (VSX; \citealt{VSX}, accessed 2022
June) is perhaps the most extensive catalogue of known variable stars.
We took all objects from the VSX labelled as CV or as NL ($n=1249$), and
then sought all for which there were available TESS SPOC light curves of
120-second cadence ($n=180$). Lomb-Scargle periodograms (LS periodogram;
\citealp{Lomb1976,Scargle1982}) of range between \mbox{0.125 --
16.000~\invday} were constructed for those systems. Periodograms were
then manually searched for the simultaneous presence of at least two
neighbouring periodicities in the region above the period gap, as well
as for one periodicity near their expected beat period. This was done to
select NLs with signatures of all $P_\text{prec}$, $P_\text{orb}$,
$P_\text{nSH}$. For most stars, long-term photometry from the All-Sky
Automated Survey for Supernovae (ASAS-SN; \citealp{asassn1,asassn2}) and
from the Catalina Sky Survey (CSS, \citealp{catalina}) was available. We
attempted to construct LS periodograms using photometry from said
surveys, but data was found to be sparse and of too long cadence to be
usable.

We report on the discovery of nSH behaviour in six poorly studied CVs.
Three of them are eclipsing systems, which enabled us to directly
determine $P_\text{orb}$. For two other systems, $P_\text{orb}$ was
identified with the use of $\varphi_\text{prec}$-dependent irradiation
of the secondary. The last CV was found to have $P_\text{prec}$ and
$P_\text{nSH}$ signatures, but not a $P_\text{orb}$ one. Our derived
value of $P_\text{orb}$ by Equation \eqref{eq:beat}, however, agrees
well with the spectroscopic measurement of \citet{Pretorius2008}. All
six objects are discussed individually in Section~\ref{sec:Results}.
During inspection, we also found five new eclipsing CVs with no
superhump behaviour. Their measured orbital periods are provided in
Table \ref{tab:ebList}, and their orbital phase curves are shown in
Figure \ref{fig:ebsummary}.

\begin{table*}
\caption{List of CVs with discovered nSHs using the methods described
in Section~\ref{sec:Analysis}. All periodicities in this table
were measured on a Lomb-Scargle periodogram of range
0.125--16~\invday~and of ten-fold oversampling. All measured
$P_\text{prec}$ in this table agree within uncertainty with the expected
values by Equation~\eqref{eq:beat} using measured $P_\text{orb}$ and
$P_\text{nSH}$. Equatorial coordinates come from Gaia DR3 and are in
the J2000 epoch.}
\label{tab:shList}
\begin{tabular}{lrrcllll}
\multicolumn{1}{c}{Name} &
\multicolumn{1}{c}{RA} &
\multicolumn{1}{c}{Dec} &
\multicolumn{1}{c}{TESS Sector} &
\multicolumn{1}{c}{$P_\text{orb}$} &
\multicolumn{1}{c}{$P_\text{nSH}$} &
\multicolumn{1}{c}{$P_\text{prec}$} &
\multicolumn{1}{c}{$|\varepsilon_-|$}
\\ \hline

HBHA 4204-09 &
$ 21\h  07\m    52\fs    24$ &
$+44\degr 05' 42\farcs 0$ &
15, 16 &
$0\fd 14128(22)$ &
$0\fd 13657(22)$ &
$4\fd 11(18)$ & 
0.0333(22)

\\[0ex]
Gaia DR3 4684361817175293440 &
$ 00\h  49\m    59\fs    93$ &
$-76\degr 08' 27\farcs 5$ &
28 &
$0\fd 15401(53)$ &
$0\fd 14750(52)$ &
$3\fd 40(27)$ &
0.0423(48)

\\[0ex]

KQ Mon &
$ 07\h  31\m    21\fs    13$ &
$-10\degr 21' 49\farcs 4$ &
34 &
$0\fd 13456(40)$ &
$0\fd 12894(38)$ &
$3\fd 12(24)$ & 
0.0418(41)

\\[0ex]
SDSS J090113.51+144704.6 &
$ 09\h  01\m    13\fs    51$ &
$+14\degr 47' 04\farcs 7$ &
44 -- 46 &
$0\fd 14631(17)$ &
$0\fd 13991(17)$ &
$3\fd 198(70)$ &
0.0437(16)

\\[0ex]
Gaia DR3 5931071148325476992 &
$ 16\h  36\m    03\fs    63$ &
$-52\degr 33' 32\farcs 6$ &
39 &
$0\fd 14827(46)$ &
$0\fd 14248(43)$ &
$3\fd 57(30)$ &
0.0391(42)

\\[0ex]
{[}PK2008{]} HalphaJ103959 &
$ 10\h  39\m    59\fs    98$ &
$-47\degr 01' 26\farcs 3$ &
36, 37 &
$0\fd 1577(2)^\dagger$ &
$0\fd 15285(29)$ &
$4\fd 94(26)$ &
0.0308(22)
\\

\hline									
\end{tabular} 							
\par\hspace{10pt}
\raggedright
$^\dagger$Orbital period measured spectroscopically by \citet{Pretorius2008}.
\end{table*}

\section{Review and results}
\label{sec:Results}

The following sections provide literature review, discussion and
interpretation of data for all CVs with discovered nSH behaviour. Each
CV system has an associated figure containing: (1) available sky-survey
data, (2) TESS photometry from sectors with prominent nSH behaviour
together with corresponding LS periodograms, (3) orbital phase plots of
data in the four $\varphi_\text{prec}$ regions discussed in Section
\ref{sec:superhump_degeneracy}. Measured periodicities of each system
are given in Table \ref{tab:shList}. All measurements agree well with
Equation \eqref{eq:beat} within uncertainty.

\subsection{HBHA 4204-09}
\begin{figure*}
    \centering
    \includegraphics{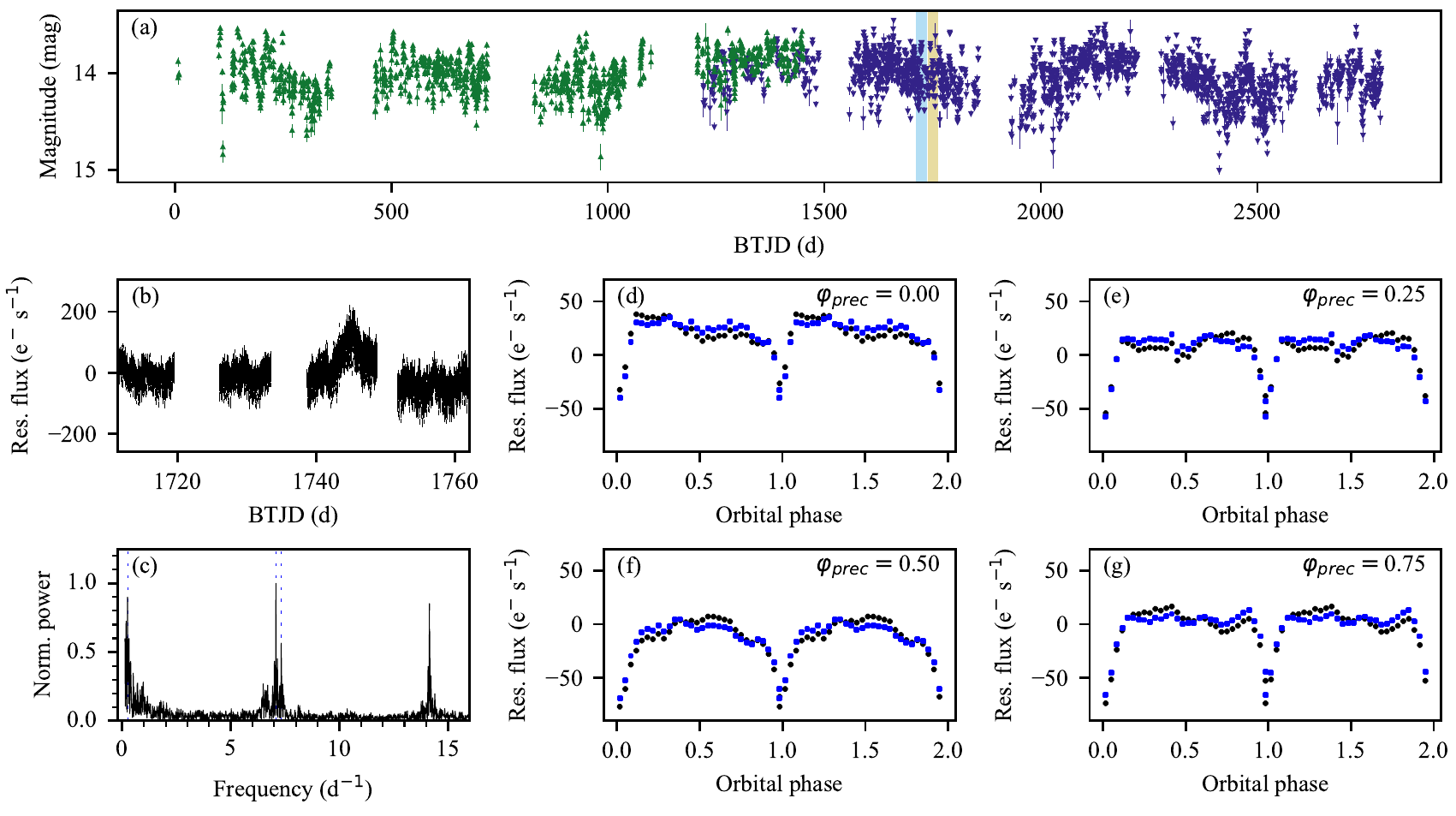}
    \caption{Photometry and analysis of \mbox{HBHA 4204-09}. (a)
    Long-term photometry from: \mbox{ASAS-SN $g$} (purple downward
    triangles), \mbox{ASAS-SN $V$} (green upward triangles). Temporal
    coverage of TESS: Sector 15 (light blue), Sector 16 (yellow). (b)
    Residual (mean-subtracted) SAP flux from TESS data. (c) Associated
    LS periodogram of (b), with indications of
    $P_\text{prec},P_\text{orb},P_\text{nSH}$ signals from Table
    \ref{tab:shList} (blue dashes). (d)--(g) Binned orbital profiles
    around disc precession phases \mbox{$\varphi_\text{prec}= 0.00,
    0.25, 0.50, 0.75$} with nSH subtraction (blue squares) and without
    one (black circles). The typical standard deviation of data in bins
    is 25 \electronpersecond. This system has a high inclination, and
    the parts of the secondary in both half-spaces are accessible to the
    observer. The secondary is expected to be the most irradiated in
    panels (d) and (f), where the observed out-of-eclipse profile is
    non-flat. In panels (e) and (g), where no irradiation of the
    secondary is expected, the out-of-eclipse profile is mostly flat.}
    \label{fig:star03summary}
\end{figure*}

HBHA 4204-09\footnote{The VSX identifier of this source is ASASSN-V
J210752.24+440542.0.} (Figure \ref{fig:star03summary}) is discovered by
ASAS-SN. It was classified as a CV by \citet{ASASSNCat12018} and by
ALeRCE \citep{ALeRCE2021} in data from the Zwicky Transient Facility
(ZTF; \citealp{ZTF}). This object is part of the ''Catalogue of Stars in
the Northern Milky Way Having H-alpha in Emission''
\citep{HaCatalogue1999}. The Gaia DR3 distance estimate is $478\pm3$~pc
and ASAS-SN photometry gives a mean brightness of $m_V = 16.19$~mag. 

We report the presence of previously unknown V-shaped eclipses in
HBHA~4204-09. Using them, we identify the periodogram peaks
corresponding to $P_\text{orb}$, $P_\text{nSH}$, $P_\text{prec}$ (Table
\ref{tab:shList}).  Aside from these periodicities, the power spectrum
contains a strong signal at $0\fd 070655(58)$, which matches
$P_\text{orb}/2$. A collection of peaks at around $0\fd 155$ is
observed, which may be indicative of a pSH signature. Additional
photometry of HBHA 4204-09 can be found in TESS Sectors~55 and 56, but
no superhumps are present in those data sources. Due to its high orbital
inclination, the near and the far half-spaces defined by the orbital
plane are comparably accessible to the observer. The portion of the
secondary in the far half-space is most irradiated at
$\varphi_\text{prec} = 0.00$, while the portion in the near half-space
is most irradiated at $\varphi_\text{prec} = 0.50$. The orbital profiles
in panels (d) and (f) of Figure~\ref{fig:star03summary} show stronger
secondary irradiation at aforementioned $\varphi_\text{prec}$, which is
expected.

\subsection{Gaia DR3 4684361817175293440}
\begin{figure*}
    \centering
    \includegraphics{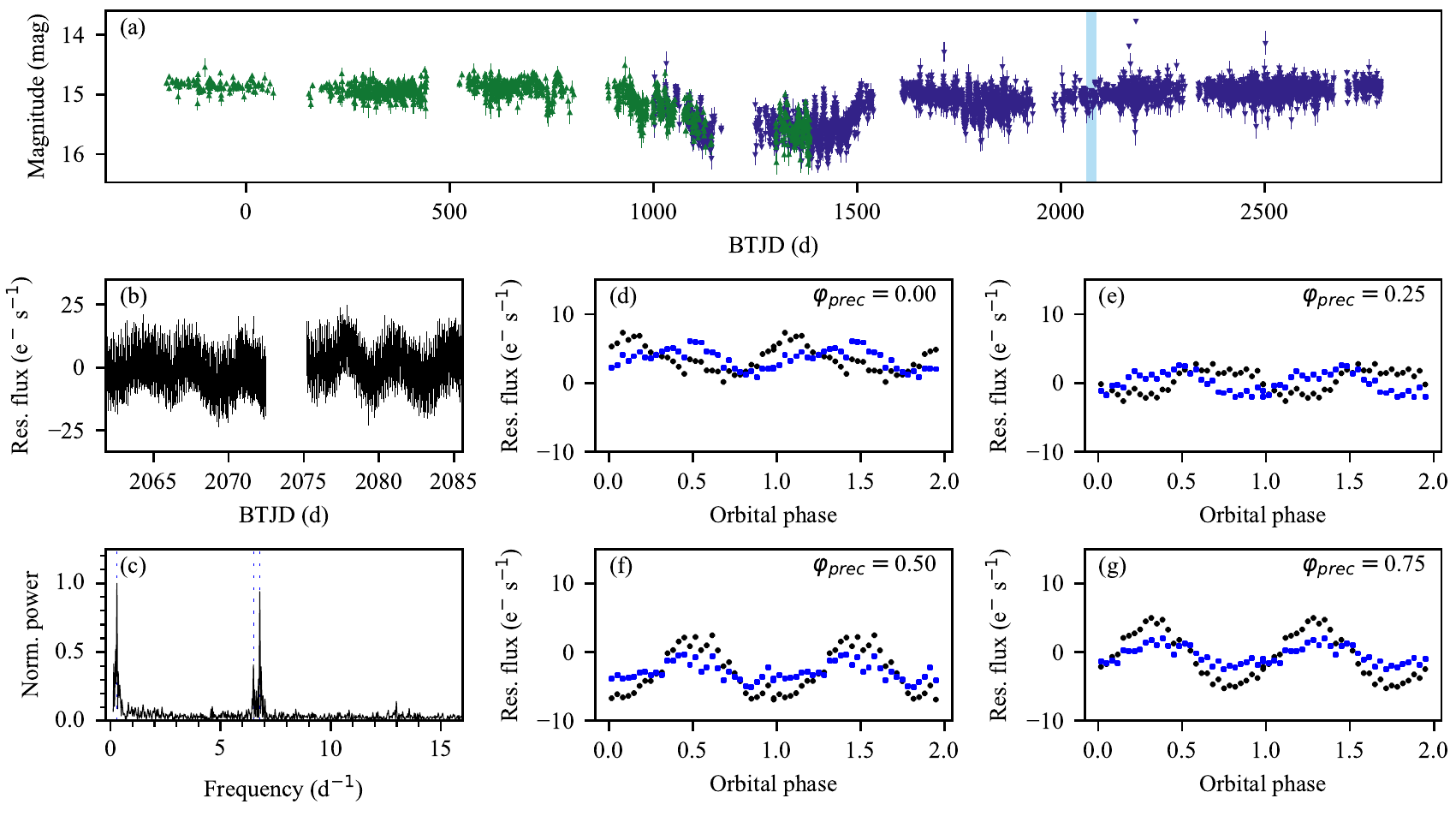}
    \caption{Photometry and analysis of \mbox{Gaia-468436}. (a)
    Long-term photometry from: \mbox{ASAS-SN $g$} (purple downward
    triangles), \mbox{ASAS-SN $V$} (green upward triangles). Temporal
    coverage of TESS: Sector 28 (light blue). (b) Residual
    (mean-subtracted) SAP flux from TESS data. (c) Associated LS
    periodogram of (b), with indications of
    $P_\text{prec},P_\text{orb},P_\text{nSH}$ signals from Table
    \ref{tab:shList} (blue dashes). (d)--(g) Binned orbital profiles
    around disc precession phases \mbox{$\varphi_\text{prec}= 0.00,
    0.25, 0.50, 0.75$} with nSH subtraction (blue squares) and without
    one (black circles). The typical standard deviation of data in bins
    is 4 \electronpersecond. The effect of variable irradiation in
    panels (d)--(g) is similar to the one observed in KIC~9406652
    \citep{Kimura2021}.}
    \label{fig:star04summary}
\end{figure*}

\begin{figure*}
    \centering
    \includegraphics[width=\textwidth]{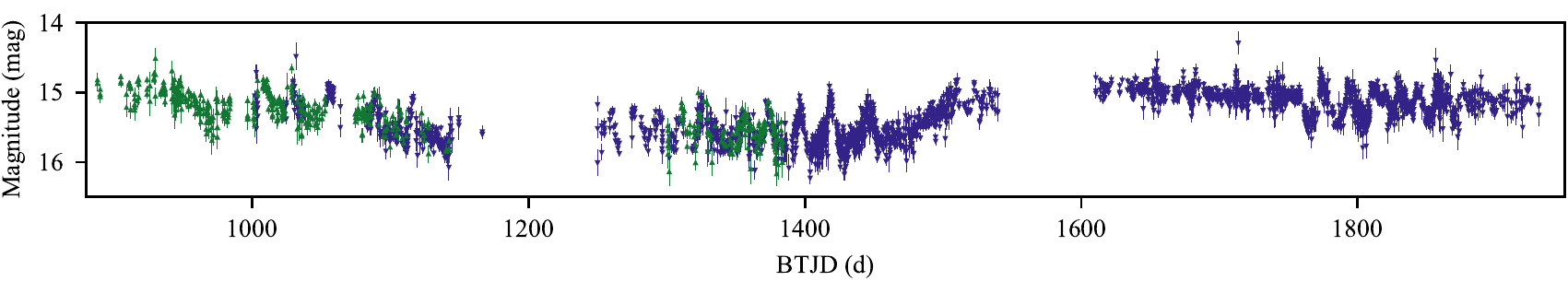}
    \caption{
    Z Cam-like episodes of Gaia-468436 in
    \mbox{ASAS-SN $g$} (blue downward triangles) and 
    \mbox{ASAS-SN $V$} (green upward triangles). 
    The Z Cam behaviour begins after a $\sim0.5$~mag fall in brightness. Emerging oscillations have a variable amplitude of $\sim 0.8$~mag and are quasi-periodic with $P \sim 20$~days. At about BTJD 1500, a brightening takes place, which is followed by a standstill at a level of 15.0~mag. At about BTJD 1760, the standstill is replaced with another oscillatory episode that has outbursts of similar period and amplitude as the former ones. No more Z~Cam episodes were observed in Gaia-468436 in this data.} 
 \label{fig:star04special}
\end{figure*}

Gaia DR3 4684361817175293440, hereinafter Gaia-468436\footnote{The VSX
identifier of this source is BMAM-V424.}
(Figure~\ref{fig:star04summary}) was discovered and classified as a NL
type CV by \citet{Bajer2019}. The Gaia DR3 distance estimate is
1062{\raisebox{0.5ex}{\tiny$^{+29}_{-30}$}}~pc. On the long-term ASAS-SN
curve, a 1-mag fall in brightness can be observed around BTJD~800 --
BTJD~1700. A panel with ASAS-SN photometry in this time period is shown
in Figure~\ref{fig:star04special}. The observed drop in brightness has a
smaller amplitude from what is expected in classic VY~Scl low states.
Quasi-cyclic variations of $P\sim20$~d resembling Z~Cam outbursts appear
after the start of the low state. The system later returns to normal
brightness and outbursts are replaced with a standstill lasting for
$\sim300$~days. This standstill is followed by another Z~Cam outburst
episode, after which no more outbursts of this type are observed. 

Our LS periodogram of Gaia-468436 shows three peaks with periods
matching Equation \eqref{eq:beat}. We interpret them as $P_\text{orb}$,
$P_\text{nSH}$,$P_\text{prec}$ in a system with a tilted precessing disc
(Section \ref{sec:superhump_degeneracy}). We find two additional peaks
at $0\fd 07372(14)$ and $0\fd 07702(15)$ that match $P_\text{nSH}/2$ and
$P_\text{orb}/2$ respectively. In
Figure~\ref{fig:star04summary}(d)--(g), it can be seen that the light
maximum of orbital-phase curves gradually shifts to earlier
$\varphi_\text{orb}$ as the disc precession cycle advances. This is
direct evidence for a retrogradely precessing tilted disc
\citep{Kimura2020}.   

\subsection{KQ~Mon}
\begin{figure*}
    \centering
    \includegraphics{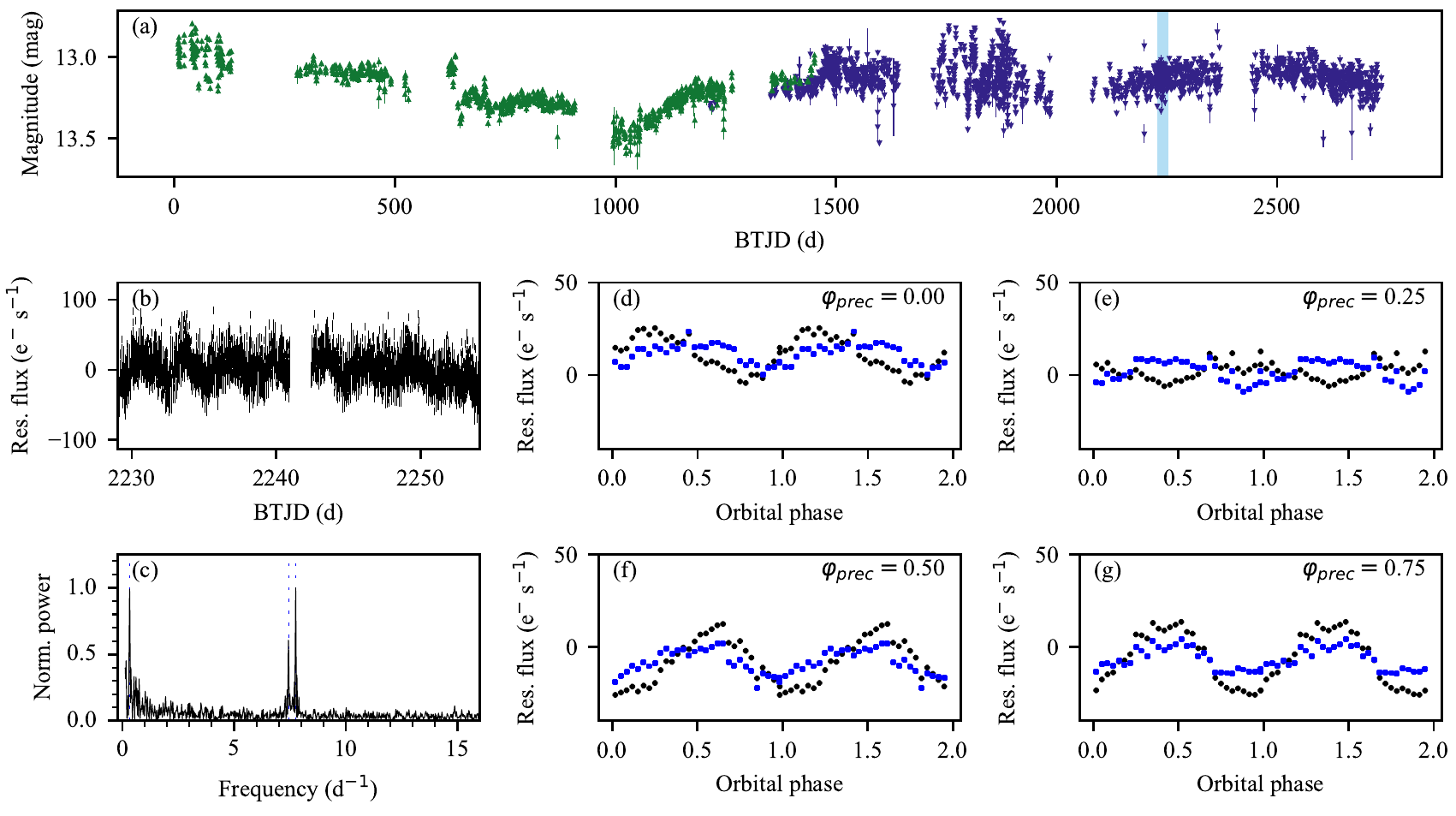}
    \caption{Photometry and analysis of \mbox{KQ Mon}. (a) Long-term
    photometry from: \mbox{ASAS-SN $g$} (purple downward triangles),
    \mbox{ASAS-SN $V$} (green upward triangles). Temporal coverage of
    TESS: Sector 34 (light blue). (b) Residual (mean-subtracted) SAP
    flux from TESS data. (c) Associated LS periodogram of (b), with
    indications of $P_\text{prec},P_\text{orb},P_\text{nSH}$ signals
    from Table \ref{tab:shList} (blue dashes). (d)--(g) Binned orbital
    profiles around disc precession phases \mbox{$\varphi_\text{prec}=
    0.00, 0.25, 0.50, 0.75$} with nSH subtraction (blue squares) and
    without one (black circles). The typical standard deviation of data
    in bins is 16 \electronpersecond. The blue and the black curves
    differ due to the significant nSH contribution to the observed
    system flux. In blue curves of different $\varphi_\text{prec}$,
    there seems to be a change of shape and amplitude near
    $\varphi_\text{orb} = 0.5$, which is expected, but could be also due
    to noise. \citet{Kimura2021} provide models for such orbital curves
    that could explain these observations.}
    \label{fig:star08summary}
\end{figure*}

KQ~Mon (Figure \ref{fig:star08summary}) was classified as a NL-type CV
by \citet{Bond1979} using low-resolution spectra in the optical. Its
orbital period was measured in \cite{Schmidtobreick2005} to be
$P_\text{orb}=0\fd 1283(17)$ by analysing two nights of time-resolved
spectroscopy. Later, \citet{Wolfe2013} examined far-ultraviolet spectra
of KQ~Mon from the International Ultraviolet Explorer. The mass of the
primary was estimated to be $M_1\sim 0.6$~$M_\odot$ with the use of
synthetic spectra. The same work argued that the primary contributes
little to the total system flux, and is overwhelmed by the flux of a
steady-state accretion disc. It was concluded that the system is located
at a distance of \mbox{144--159~pc}, with an inclination of $i\leq
60\degr$ and an accretion rate in the order of $\dot{M}\sim
10^{-9}$~$M_\odot$~yr$^{-1}$. The Gaia DR3 distance is
\mbox{628$\pm8$~pc}, which disagrees with their estimates.

Our measured value for $P_\text{nSH}$ matches the $P_\text{orb}$ given
in \citet{Schmidtobreick2005}. What we measure as
\mbox{$P_\text{orb}=0\fd 13456(40)$} would have corresponded to a pSH
signal in their interpretation. But then, no other signals in the
periodogram would have been expected. We, however, measure a strong
third signal at $3\fd 12(24)$, which is self-consistent with the other
two by Equation \eqref{eq:beat}. In addition, we observe a
$\varphi_\text{prec}$-dependent amplitude of the orbital phase
curve, which could be explained by a varying irradiation of the
secondary. In Figure~3 of \citet{Schmidtobreick2005}, a strong aliasing
pattern can be seen. The authors chose an orbital period $P_\text{orb}$
among four possible signals, two of which agree with our measurements of
$P_\text{orb},P_\text{nSH}$. With all this in mind, we think that the
correct value of $P_\text{orb}$ is $0\fd 13456(40)$, and that there is
presence of a tilted accretion disc in this system.

\subsection{SDSS J090113.51+144704.6}
\begin{figure*}
    \centering
    \includegraphics{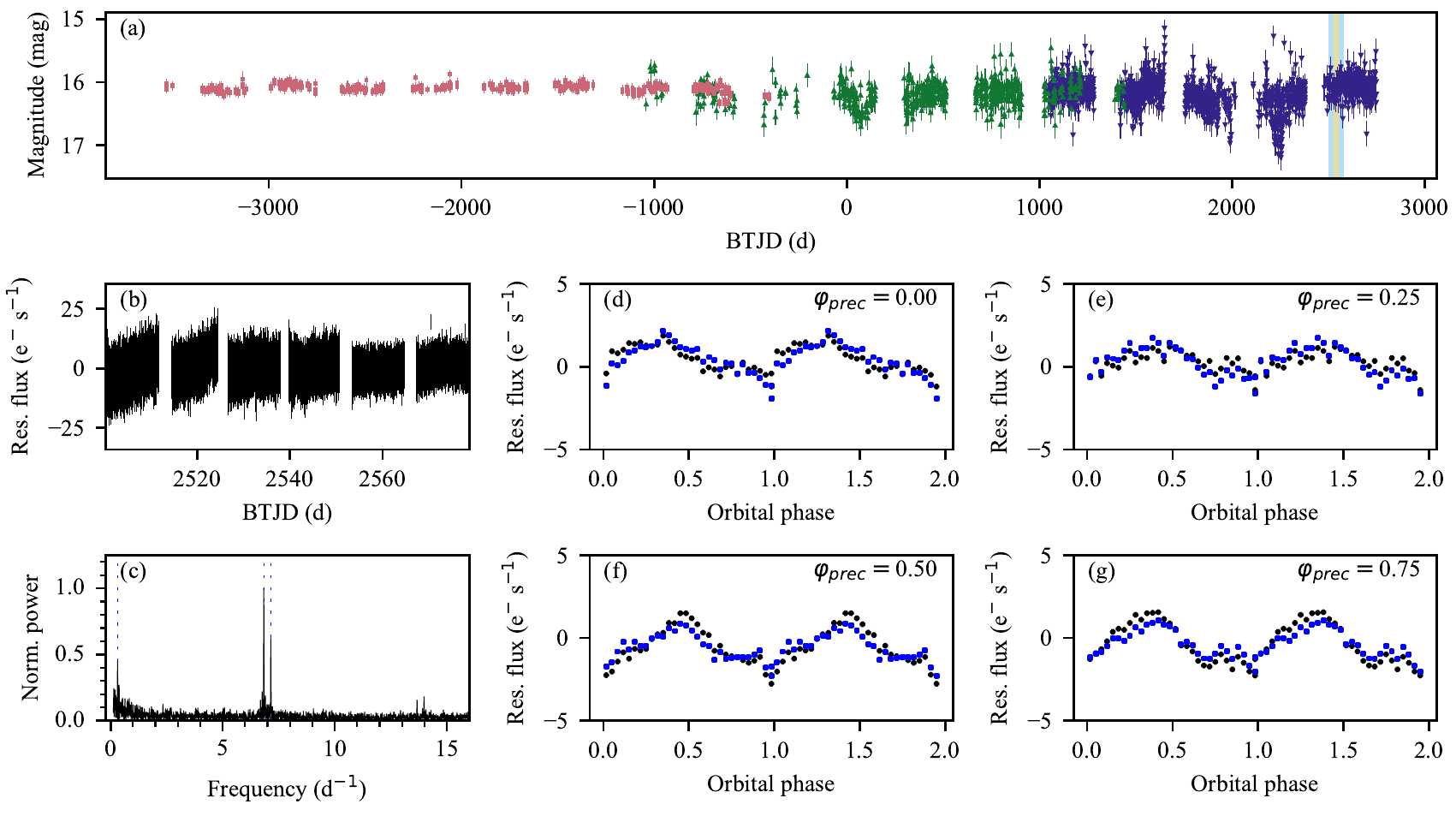}
    \caption{Photometry and analysis of \mbox{SDSS-090113}. (a)
    Long-term photometry from: ASAS-SN $g$ (purple downward triangles),
    ASAS-SN $V$ (green upward triangles), Catalina  $V$ (pink squares).
    Temporal coverage of TESS: Sector 44 (light blue), Sector 45
    (yellow), Sector 46 (light blue). (b) Residual (mean-subtracted) SAP
    flux from TESS data. (c) Associated LS periodogram of (b), with
    indications of $P_\text{prec},P_\text{orb},P_\text{nSH}$ signals
    from Table \ref{tab:shList} (blue dashes). Data from (b) was
    smoothed by a fourth-order Savitzky-Golay filter of window size 10 d
    before constructing the periodogram. This was done solely for the
    sake of clear identification of $P_\text{prec}$ by the reader, and
    not for periodicity measurements. (d)--(g) Binned orbital profiles
    around disc precession phases \mbox{$\varphi_\text{prec}= 0.00,
    0.25, 0.50, 0.75$} with nSH subtraction (blue squares) and without
    one (black circles). The typical standard deviation of data in bins
    is 4 \electronpersecond. SDSS-090113 appears to have shallow grazing
    eclipses that are barely detectable for some $\varphi_\text{prec}$.
    Observed eclipses vary in depth and width for different
    $\varphi_\text{prec}$. This could be explained by a secondary that
    partially covers the tilted disc only when the projected area of the
    disc is large.}
    \label{fig:star13summary}
\end{figure*}

SDSS J090113.51+144704.6, hereinafter SDSS-090113 (Figure
\ref{fig:star13summary}) first appeared in \citet{Szkody2009} where it
was classified as a CV due to accretion disc features in its spectrum.
This system was included in the catalogue of bright WDs of
\cite{Raddi2017}. Gaia DR3 estimated the distance to SDSS-090113 to be
1482{\raisebox{0.5ex}{\tiny$^{+100}_{-116}$}}~pc. Later,
\citet{Mosenlechner2022} included this system in their time-series
analysis study of subdwarf A-type stars using Kepler~K2 data. They
discovered a periodicity of $0\fd 146$, which was suggested to be the
orbital period $P_\text{orb}$. 

\begin{figure*}
    \centering
    \includegraphics[width=\textwidth]{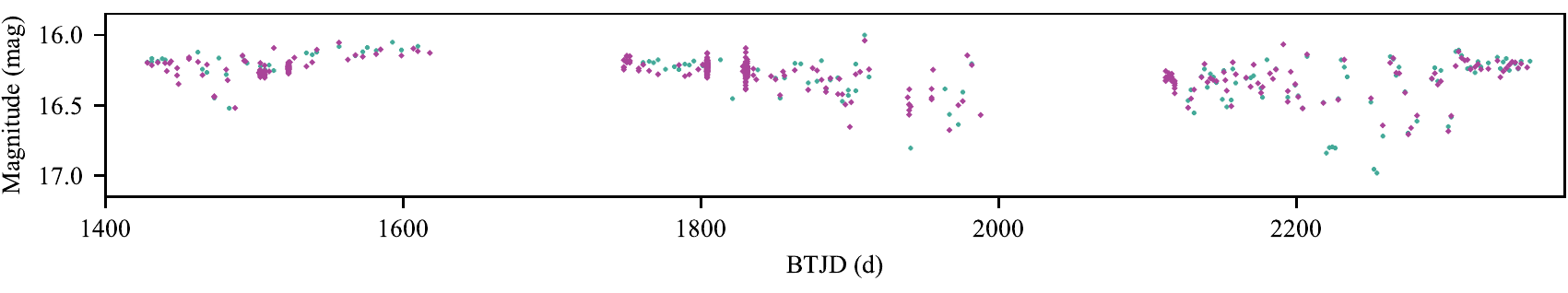}
    \caption{IW~And episodes of SDSS-090113 in \mbox{ZTF $g$} (teal
    circles) and \mbox{ZTF $r$} (magenta diamonds). Three seasons of
    photometry are shown. The first season starts with oscillations that
    are terminated by brightening. This is one of the defining features
    of the IW~And phenomenon \citep{Kato2019,KatoASAShrana2022}. The
    second season shows the beginning of a new oscillatory episode that
    is variable in amplitude. The episode continues in the third season
    and abruptly ends at BTJD 2315.}
 \label{fig:star13special}
\end{figure*}

SDSS-090113 has no recorded low states and its brightness varies around
$m_V=16.2$~mag. Between BTJD~1600 and BTJD~2300, we recognise an episode
of anomalous Z~Cam-type outbursts repeating once about every 25 days
(Figure \ref{fig:star13special}). These outbursts begin after a
brightening, which is one of the defining features of the
IW~And-phenomenon systems \citep{Kato2019}. This can be explained by a
tilted disc that causes the accretion stream to enter inner disc
regions, and thus to disrupt the accretion cycle. In this new type of
accretion, the inner disc is in a hot state, while the outer disc
repeats outbursts \citep{KimuraIWAnd2020}.

Figure~\ref{fig:star13summary}(d)--(g) shows what seems to be the
presence of grazing eclipses in the orbital curve of SDSS-090113. They
vary in depth and width, and for some phases of $P_\text{prec}$ they
disappear completely, similar to ES~Dra \citep{Kato2022}. Through them,
we identify $P_\text{orb}, P_\text{nSH}, P_\text{prec}$ (Table
\ref{tab:shList}). Two additional peaks are found at $0\fd 071531(52)$
and $0\fd 073161(40)$ that match $P_\text{nSH}/2$ and $P_\text{orb}/2$
respectively.

\subsection{Gaia DR3 5931071148325476992}\label{sec:gaia_crowded}
\begin{figure*}
    \centering
    \includegraphics{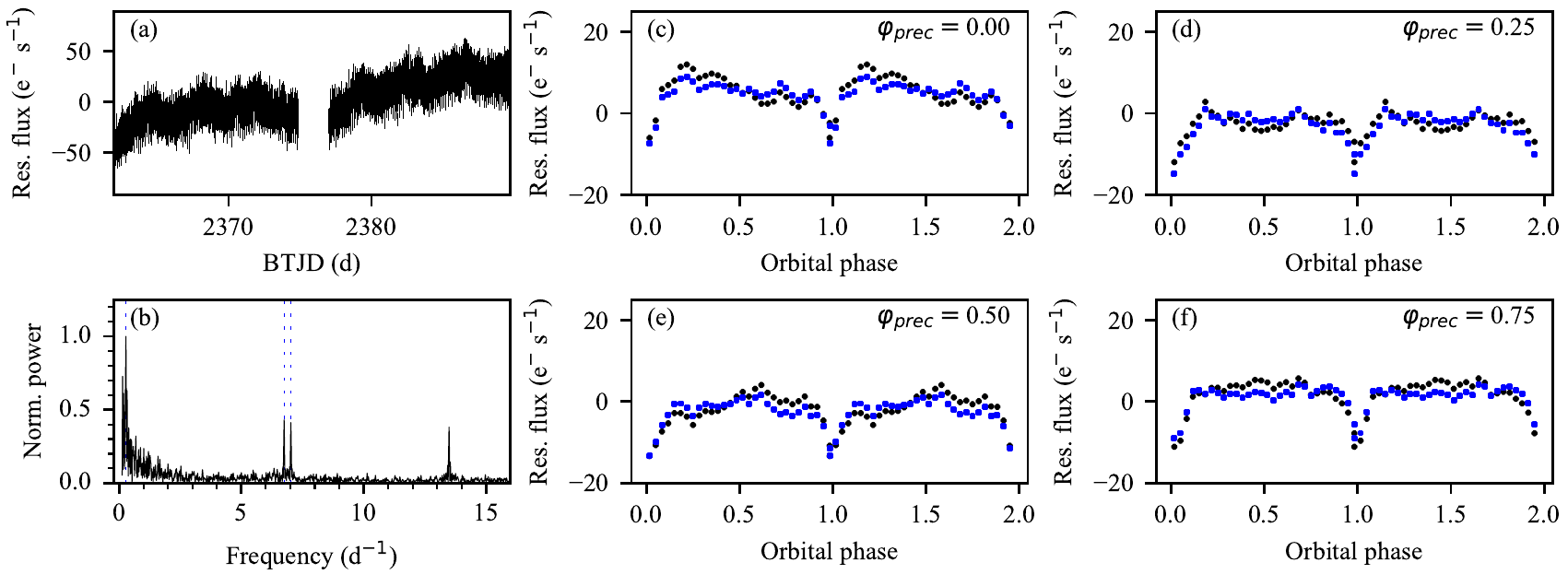}
    \caption{Photometry and analysis of \mbox{Gaia-593107}. (a) Residual
    (mean-subtracted) SAP flux from TESS data. (b) Associated LS
    periodogram of (a), with indications of
    $P_\text{prec},P_\text{orb},P_\text{nSH}$ signals from Table
    \ref{tab:shList} (blue dashes). (c)--(f) Binned orbital profiles
    around disc precession phases \mbox{$\varphi_\text{prec}= 0.00,
    0.25, 0.50, 0.75$} with nSH subtraction (blue squares) and without
    one (black circles). The typical standard deviation of data in bins
    is 8 \electronpersecond. Similar to the other eclipsing binaries in
    our sample, the secondary is expected to be the most irradiated in
    panels (c) and (e), where the observed out-of-eclipse profile is
    non-flat. In panels (d) and (f), these profiles are mostly flat, and
    the system brightness does not seem to increase near
    $\varphi_\text{orb} = 0.5$.}
    \label{fig:star14summary}
\end{figure*}

Gaia~DR3~5931071148325476992, hereinafter Gaia-593107\footnote{The VSX
identifier of this source is USNO-A2.0 0300-28957281.} (Figure
\ref{fig:star14summary}) is a poorly studied CV that was discovered in
plates by \citet{Prestgard2020} from the Digitized Sky
Survey\footnote{ESO Online Digitized Sky Survey:
\url{http://archive.eso.org/dss/dss} (accessed 2022 October).} and the
SuperCOSMOS H$\alpha$ survey \citep{SHS2005}. The NOMAD catalogue
\citep{NOMAD2004} gives an apparent magnitude of $m_V=16$~mag. No
\mbox{ASAS-SN} photometry is available for this system. Its TESS brightness
reads $m_\text{TESS}=16.02$ mag. There is an X-ray source
\mbox{(1RXS~J163605.9-523335)} at a distance of 20.8 arcsec, which is likely
associated with Gaia-593107. In addition, we find two bright sources of
brightness $m_\text{TESS}=13.44$ and $m_\text{TESS}=15.16$ mag in the
aperture mask, that are expected to severely contaminate the light
curve. This issue, however, is resolved by the apparent variability of
Gaia-593107 in the discovery
images\footnote{\url{https://www.aavso.org/vsx_docs/1544030/3344/USNO-A2.0\%200300-28957281.png}}
of \citeauthor{Prestgard2020}, on the basis of which we attribute the
tilted-disc behaviour to this specific system.

Our analysis of TESS light curves shows Gaia-593107 to be an eclipsing
variable with an orbital period of $P_\text{orb}=0\fd 14248(43)$. A peak
at $P_\text{orb}/2$ is present as well. This allows us to locate
$P_\text{orb}, P_\text{nSH}, P_\text{prec}$ (Table \ref{tab:shList}). A
change in eclipse depth is observed in different phases of the
determined $\varphi_\text{prec}$.  

\subsection{[PK2008] HalphaJ103959}
\begin{figure}
    \centering
    \includegraphics{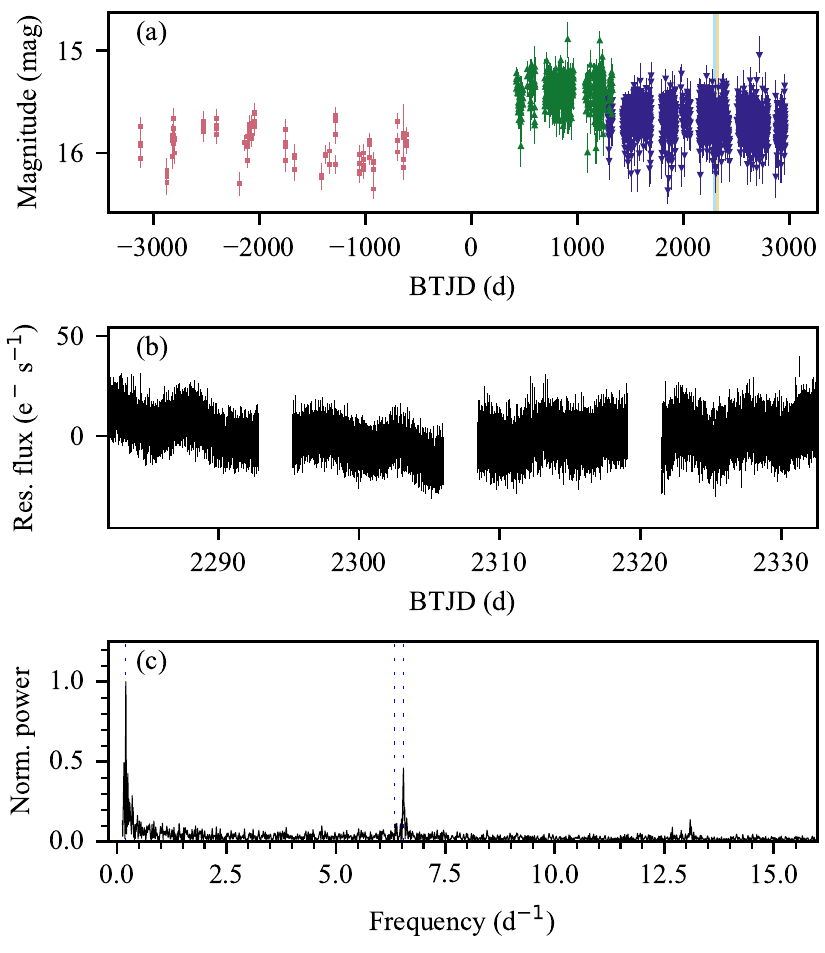}
    \caption{Photometry and analysis of \mbox{PK-103959}. (a) Long-term
    photometry from: ASAS-SN $g$ (purple downward triangles), ASAS-SN $V$
    (green upward triangles), Catalina  $V$ (pink squares). Temporal
    coverage of TESS: Sector 36 (light blue), Sector 37 (yellow). (b)
    Residual (mean-subtracted) SAP flux from TESS data. (c) Associated
    LS periodogram of (b), with indications of
    $P_\text{prec},P_\text{orb},P_\text{nSH}$ signals from Table
    \ref{tab:shList} (blue dashes).}
    \label{fig:star15summary}
\end{figure}

[PK2008] HalphaJ103959, hereinafter PK-103959  (Figure
\ref{fig:star15summary}) was classified as a CV in
\citet{Pretorius2008}, where spectroscopic and photometric analyses of
the system were carried out. The orbital period of PK-103959 was
measured to be $P_\text{orb}=0\fd 1577(2)$ in the same work. Catalina
and ASAS-SN photometry has a mean brightness of $m_V=$15.7~mag, with no
low states. A gradual increase in brightness can be seen in the period
between -1000 and 2000~BTJD.

We find signatures of $P_\text{nSH}$ and $P_\text{prec}$ (Table
\ref{tab:shList}), but no peaks at the $P_\text{orb}$ by
\citeauthor{Pretorius2008}. However, there are two other visible peaks
at $0\fd 07885(14)$ and $0\fd 07639(13)$, which correspond to
$P_\text{orb}/2$ by \citeauthor{Pretorius2008} and $P_\text{nSH}/2$
respectively. 

\section{Discussion}\label{sec:Discussion}

\subsection{Mass-ratio estimates}
\label{sec:Discussion_massratio}
\begin{figure*}
\centering
\includegraphics{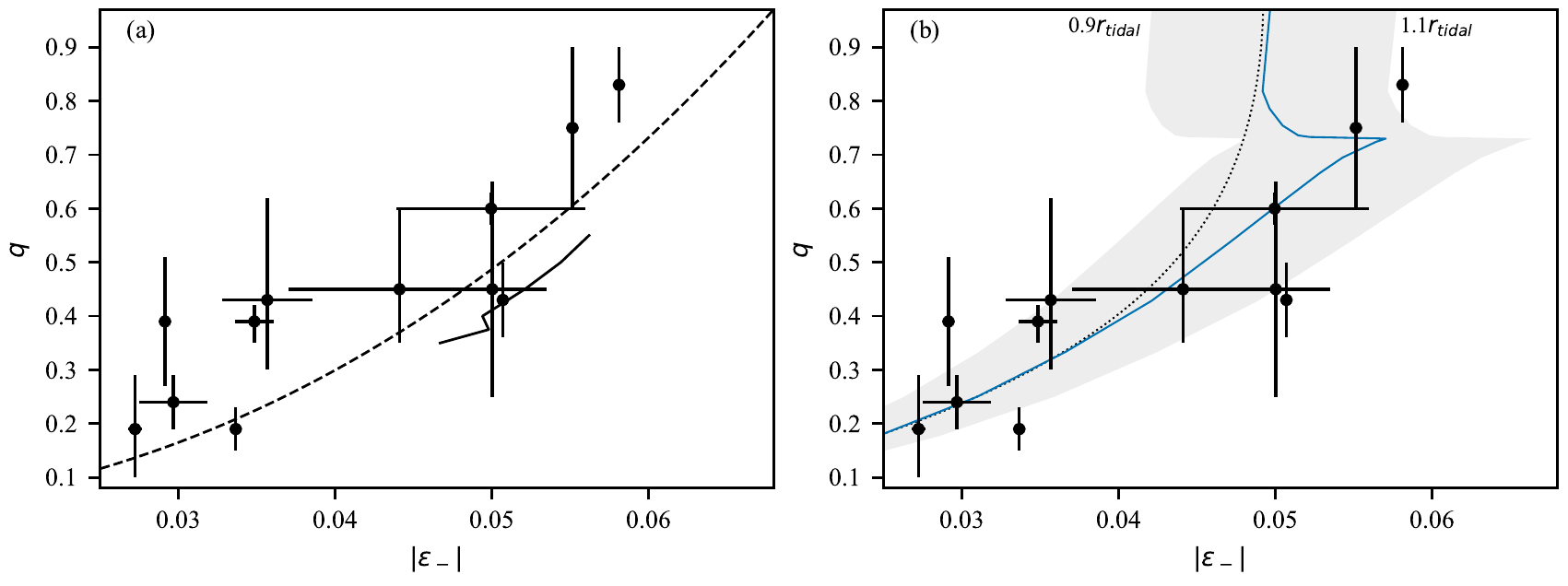}
\caption{Different $q(\varepsilon_-)$ relations against NL variables
with $\varepsilon_-,q$ measurements from Table \ref{tab:stars_q_e}. (a)
Solid line: \citet{Montgomery2009}, Dashed line: \citet{Wood2009}. Both
relations underestimate $q(\varepsilon_-)$ for given measurements. (b)
Dotted line: Equation \eqref{eq:Warner1995final}, derived from the
approximation by \citet{Warner1995} used in Equation
\eqref{eq:epsilon-q_relation}. This fails to accurately describe
\citet{Paczynski1977} in $q$ regimes near 0.7. Blue curve: a computed
$q(\varepsilon_-)$ relation from our treatment in Section
\ref{sec:Discussion_massratio} that makes use of \citet{Paczynski1977}.
Shaded region: solutions between $R_d=$0.9--1.1$r_\text{tidal}$. All
measurements belong to this region within uncertainty.}
\label{fig:qeps}
\end{figure*}

By using smoothed particle hydrodynamic (SPH) simulations of tilted
accretion discs, \citet{Wood2009} found that the relation between the
mass ratio and the nSH deficit is well-represented by
\begin{equation}
\label{eq:Wood2009}
q(\varepsilon_-) =
-0.192|\varepsilon_-|^{0.5}
+10.37|\varepsilon_-|
-99.83|\varepsilon_-|^{1.5}
+451.1|\varepsilon_-|^2.
\end{equation}
This result has been supported by other works using related SPH
simulations \citep{Montgomery2009,Thomas2015}. To compare Equation
\eqref{eq:Wood2009} with observations, we searched for NL objects in
literature for which superhump deficits and mass ratios were measured
independently from one another. Our reasoning is that NLs share three
main similarities with our discovered CVs: (1) their $P_\text{orb}$ are
of the same order, (2) they have steady, hot and luminous discs, (3)
samples of both populations exhibit VY-Scl behaviour. Using the sample
of NLs with nSH signatures, given in \citet{Bruch2023}, we were able to
find twelve such objects, which we list in Table~\ref{tab:stars_q_e}.
Figure~\ref{fig:qeps}(a) compares their measurements with the
$q(\varepsilon_-)$ relations provided by \citet{Montgomery2009} and
\citet{Wood2009}. We share the concern that both works underestimate
$q(\varepsilon_-)$ with respect to past measurements in literature.

There exists a different approach that could estimate mass ratios using
nSHs. By making some assumptions, a $q(\varepsilon_-)$ relation can be
derived in the following manner. Through linear perturbation theory,
\citet{Papaloizou1995} derived the precession rate $\omega_\text{prec}$
of a differentially rotating fluid disc with a mass profile $\Sigma(r)$:
\begin{equation}
    \omega_\text{prec}  = 
    -\dfrac{3}{4}\dfrac{GM_2}{a^3}
    \dfrac{\int\Sigma r^3 \dd r}
    {\int \Sigma \Omega r^3 \dd r} \cos\theta,
\end{equation}
where $a$ is the orbital separation, $\Omega(r) = \sqrt{GM_1/r^3}$ is
the Keplerian angular velocity profile of the disc, and $\theta$ is the
disc tilt with respect to the orbital plane. For a power-law mass
profile $\Sigma(r)\propto r^n$, \citet{OsakiKato2013} derived that
\begin{equation}
\label{eq:OsakiKato2013}
\frac{\nu_\text{prec}}
{\nu_\text{orb}} =
-\frac{3}{4}
\frac{2.5+n}{4+n}
\frac{q}{\sqrt{1+q}}
\left(\frac{R_d}
{a}
\right)^{1.5}\cos\theta,
\end{equation}
where $\nu=\omega/2\pi$ and $R_d$ is the disc radius.\footnote{We note
that precession is retrograde, which implies
$\omega_\text{prec},\nu_\text{prec}<0$.} Using Equations
\eqref{eq:beat}, \eqref{eq:epsilon-} and a mass profile of a
steady-state disc given by $n=-0.75$ \citep{ShakuraSunyaev1973},
Equation \eqref{eq:OsakiKato2013} can be reduced to
\begin{equation}
\label{eq:epsilon-q_relation}
\frac{\varepsilon_-}{1+\varepsilon_-}
=
-\frac{21}{52}
\frac{q}{\sqrt{1+q}}
\left(\frac{R_d}
{a}
\right)^{1.5}\cos\theta.
\end{equation}
A similar derivation can be found in
\citet{Montgomery_earthsunmoon2009}. This shows that $\varepsilon_-$
depends on three parameters: the mass ratio $q$, the disc tilt $\theta$
and the fractional disc radius $R_d/a$. The third can be reasoned to be
a function of $q$ as follows. Suppose that in our systems with
discovered nSHs, accretion discs are in steady state most of the time.
Then, $R_d$ approaches the tidal truncation radius $r_\text{tidal}$.
\citet[Table 1]{Paczynski1977} provided a functional dependence
$r_\text{tidal}(q)$. Later, \citet{Warner1995} proposed the
approximation
\begin{equation}
\label{eq:Warner1995}
r_\text{tidal}=\frac{0.60a}{1+q}
\end{equation}
for $0.03<q<1$. This is a good approximation in all regions but near
$q=0.7$, where $r_\text{tidal}(q)$ is underestimated. Using it would
reduce Equation \eqref{eq:epsilon-q_relation} to 
\begin{equation}
\label{eq:Warner1995final}
\frac{\varepsilon_-}{1+\varepsilon_-}=
-\frac{0.188q}{(1+q)^2}\cos\theta,
\end{equation}
which does not describe well observational data for $q>0.4$ (see dotted
line in Figure \ref{fig:qeps}(b)). Because of this, we do not use
Equation~\eqref{eq:Warner1995final}. Instead, we linearly interpolate
between data given in \citet[Table 1]{Paczynski1977} in order to
evaluate $R_d/a$ in Equation \eqref{eq:epsilon-q_relation}.

The other independent variable in Equation \eqref{eq:epsilon-q_relation}
is the disc tilt $\theta$. \citet{Smak2009} predicts that disc tilts
should not exceed $\theta_\text{max} = 7\degr$ for CVs. In their
photometric analysis of KIC~9406652, \citet{Kimura2020} concluded that
$\theta$ varies between 0--3$\degr$ over the course of 1500 days. Such
range of $\theta$ allows the assumption $\cos\theta\simeq 1$, which is
accurate to within one per cent. This motivates us to compute a
$q(\varepsilon_-)$ curve for \mbox{$\cos\theta=1$} and compare it
against measurements of Table~\ref{tab:stars_q_e} objects
(Figure~\ref{fig:qeps}(b); Table~\ref{tab:lookup_table}). The
$q(\varepsilon_-)$ relation becomes two-fold degenerate in $q$ from
about $|\varepsilon_-| > 0.048$.

\begin{table*}
\caption{
A list of NLs with independently measured superhump deficit
$\varepsilon_{-}$ and mass ratio $q$. Equatorial coordinates come from
Gaia DR3 and are in the J2000 epoch. Other references:
($a$) \citet{Gies2013};
($b$) \citet{Africano1978};
($c$) \citet{Smak2019};
($d$) \citet{Subebekova2020};
($e$) \citet{Skillman1995};
($f$) \citet{Bruch2022};
($g$) \citet{Rodriguez-Gil2020};
($h$) \citet{Kozhevnikov2007};
($i$) \citet{Araujo-Betancor2003};
($j$) \citet{Boyd2017};
($k$) \citet{XiaoanWu2002};
($l$) \citet{Huber1998};
($m$) \citet{Taylor1998};
($n$) \citet{HoardSzkody1997};
($o$) \citet{Patterson1999};
($p$) \citet{Arenas2000};
($q$) \citet{PetersThorstensen2006};
($r$) \citet{Patterson1997};
($s$) \citet{Neustroev2011};
($t$) \citet{deMiguel2016};
($u$) \citet{Szkody1993};
($v$) \citet{Bruch2023};
($w$) \citet{Gulsecen2009};
($x$) \citet{Hellier1993};
($y$) \citet{Bruch2022}.
}

\label{tab:stars_q_e}
\begin{tabular}{lrrllll}
\multicolumn{1}{c}{Name} &
\multicolumn{1}{c}{RA} &
\multicolumn{1}{c}{Dec} &
\multicolumn{1}{c}{$q$} &
\multicolumn{1}{c}{$P_\text{orb}$} &
\multicolumn{1}{c}{$P_\text{nSH}$} &
\multicolumn{1}{c}{$|\varepsilon_{-}|$}

\\\hline

KIC 9406652 &
$ 19\h 31\m    29\fs    15$ &
$+45\degr 59' 06\farcs 1$ &
$^a0.83\pm0.07$ &
$^a0.25450(2)$ &
$^a0.23971(2)$ &
$0.0581(1)$ \\[0.5ex]

RW Tri &
$ 02\h 25\m    36\fs    16$ &
$+28\degr 05' 50\farcs 9$ &
$^d0.60\pm0.03$ &
$^b0.23188324(4)$ &
$^c0.2203(14)$ &
$0.050(6)$ \\[0.5ex]

MV Lyr &
$ 19\h 07\m    16\fs    29$ &
$+44\degr 01' 07\farcs 9$ &
$^e0.43^{+0.19}_{-0.13}$ &
$^e0.1329(4)$ &
$^f0.12816(1)$ &
$0.036(3)$ \\[0.5ex]

KR Aur &
$ 06\h 15\m    43\fs    92$ &
$+28\degr 35' 08\farcs 6$ &
$^g0.39^{+0.03}_{-0.04}$ &
$^g0.16277164(5)$ &
$^h0.1571(2)$ &
$0.035(1)$ \\[0.5ex]

DW UMa &
$ 10\h 33\m    52\fs    88$ &
$+58\degr 46' 54\farcs 7$ &
$^i0.39\pm 0.12$ &
$^i0.136606499(3)$ &
$^j0.132626(9)$ &
$0.02914(7)$ \\[0.5ex]

TT Ari &
$ 02\h 06\m    53\fs    08$ &
$+15\degr 17' 41\farcs 9$ &
$^k0.19\pm0.04$ &
$^k0.1375504(17)$ &
$^f0.132921(2)$ &
$0.03366(2)$ \\[0.5ex]

V592 Cas &
$ 00\h 20\m    52\fs    22$ &
$+55\degr 42' 16\farcs 2$ &
$^l0.19^{+0.10}_{-0.09}$ &
$^m0.115063(1)$ &
$^m0.11193(5)$ &
$0.0272(4)$ \\[0.5ex]

BH Lyn &
$ 08\h 22\m    36\fs    05$ &
$+51\degr 05' 24\farcs 6$ &
$^n0.45^{+0.15}_{-0.10}$ &
$^n0.15587520(5)$ &
$^o0.1490(11)$ &
$0.044(7)$ \\[0.5ex]

V603 Aql &
$ 18\h 48\m    54\fs    64$ &
$+00\degr 35' 02\farcs 9$ &
$^p0.24\pm0.05$ &
$^q0.13820103(8)$ &
$^r0.1341(3)$ &
$0.030(2)$ \\[0.5ex]

UX UMa &
$ 13\h 36\m   40\fs   95$ &
$+51\degr 54' 49\farcs 4$ &
$^s0.43\pm0.07$ &
$^t0.19667118(19)$ &
$^t0.186700(11)$ &
$0.05070(6)$ \\[0.5ex]

AY Psc &
$ 01\h 36\m   55\fs   46$ &
$+07\degr 16' 29\farcs 3$ &
$^u0.45\pm0.2$ &
$^v0.217320654(4)$ &
$^w0.20645(75)$ &
$0.050(3)$ \\[0.5ex]

TV Col &
$ 05\h 29\m   25\fs   53$ &
$-32\degr 49' 03\farcs 9$ &
$^x0.75\pm0.15$ &
$^y0.22860010(2)$ &
$^y0.215995(1)$ &
$0.055140(4)$

\end{tabular}
\end{table*}

There are several systems that lie far from our computed curve. However,
they would all agree with a $q(\varepsilon_-)$ relation where $R_d$ is
between 0.9--1.1$r_\text{tidal}$ (also in Figure~\ref{fig:qeps}(b)). It
becomes apparent that the curve is much more sensitive to changes in
$R_d$ than to changes in $\theta$. On this basis, we compute three
curves for different
\mbox{$R_d=0.9r_\text{tidal},1.0r_\text{tidal},1.1r_\text{tidal}$} and
then we graphically solve $q$ for our measured $\varepsilon_-$. Our
mass-ratio estimates are listed in Table~\ref{tab:shmassratios}.

Systems with smaller $|\varepsilon_-|$ tend to be better constrained in
$q$. For example, PK-103959 has the lowest $|\varepsilon_-|=0.0308(22)$
in our sample, and its mass ratio shows little variation for different
values of $R_\text{d}$. Conversely, SDSS-090113 has the largest measured
$|\varepsilon_-|$ in our sample. For different $R_\text{d}$, its $q$
estimates differ significantly with respect to statistical uncertainty.

\begin{table}
\centering
\caption{ Estimates of the mass ratio $q$ of Table \ref{tab:shList}
systems for disc radii $R_d=0.9r_\text{tidal}, 1.0r_\text{tidal}$, $1.1
r_\text{tidal}$ from \citet{Paczynski1977}, using the methods described
in Section~\ref{sec:Discussion_massratio}. Upper bounds of $q$ entering
the region of two-fold degeneracy are labeled with an asterisk.  
}
\label{tab:shmassratios}
\begin{tabular}{clll}
\multirow{2}{*}{Name} & \multicolumn{3}{c}{$q(\varepsilon_-)$} \\[0.5ex]
\cline{2-4}\\[-1.5ex]
 & $0.9 r_\text{tidal}$ & $1.0 r_\text{tidal}$ & $1.1 r_\text{tidal}$
\\ \hline

\multicolumn{1}{l}{HBHA 4204-09} &
0.38{\raisebox{0.5ex}{\tiny$^{+0.05}_{-0.05}$}} &
0.28{\raisebox{0.5ex}{\tiny$^{+0.03}_{-0.03}$}} &
0.22{\raisebox{0.5ex}{\tiny$^{+0.02}_{-0.02}$}} 
\\[0.2ex]

\multicolumn{1}{l}{Gaia-468436} &
0.60{\raisebox{0.5ex}{\tiny$^{+0.09^{*}}_{-0.13}$}} &
0.43{\raisebox{0.5ex}{\tiny$^{+0.10}_{-0.08}$}} &
0.33{\raisebox{0.5ex}{\tiny$^{+0.07}_{-0.06}$}} 
\\[0.2ex]

\multicolumn{1}{l}{KQ Mon} &
0.59{\raisebox{0.5ex}{\tiny$^{+0.11^{*}}_{-0.11}$}} &
0.42{\raisebox{0.5ex}{\tiny$^{+0.09}_{-0.07}$}} &
0.32{\raisebox{0.5ex}{\tiny$^{+0.06}_{-0.05}$}} 
\\[0.2ex]

\multicolumn{1}{l}{SDSS-090113} &
0.64{\raisebox{0.5ex}{\tiny$^{+0.04^{*}}_{-0.04}$}} &
0.46{\raisebox{0.5ex}{\tiny$^{+0.03}_{-0.03}$}} &
0.35{\raisebox{0.5ex}{\tiny$^{+0.02}_{-0.02}$}} 
\\[0.2ex]

\multicolumn{1}{l}{Gaia-593107} &
0.52{\raisebox{0.5ex}{\tiny$^{+0.11}_{-0.10}$}} &
0.38{\raisebox{0.5ex}{\tiny$^{+0.08}_{-0.07}$}} &
0.29{\raisebox{0.5ex}{\tiny$^{+0.06}_{-0.05}$}} 
\\[0.2ex]

\multicolumn{1}{l}{PK-103959}  &
0.33{\raisebox{0.5ex}{\tiny$^{+0.04}_{-0.04}$}} &
0.25{\raisebox{0.5ex}{\tiny$^{+0.03}_{-0.03}$}} &
0.20{\raisebox{0.5ex}{\tiny$^{+0.02}_{-0.02}$}} 
\end{tabular} 							
\end{table}
 
\subsection{Orbital inclination estimates}\label{sec:incl+tilt}
The eclipse width at half depth $\Delta \varphi_\text{orb}$ is a
reasonable measure of the primary-eclipse duration in the assumption of
an axisymmetric disc \citep[Section~2.6.2]{Warner1995}. This duration
can be used to determine the orbital inclination $i$ by the
relationship
\begin{equation}
    \sin^2 i \approx \dfrac{
    1-R_L(2)^2
    }{\cos^2(2\pi\varphi_p)},
    \label{eq:HorneWarner}
\end{equation}
where $R_{L}(2)$ is the radius of the secondary star, and $\pm
\varphi_p$ are the phases of mid-immersion and mid-emergence of the
primary star \citep{Horne1985,Warner1995}.\footnote{From here,
$\varphi_p\equiv\Delta\varphi_\text{orb}/2$.} The Roche-lobe radius
approximation by \citet{Eggleton1983}
\begin{equation}
    R_L(2) = \frac{
    0.49q^{2/3}
    }{
    0.6q^{2/3}+\ln(1+q^{1/3})
    }
\end{equation}
can be used to substitute $R_L(2)$ in Equation \eqref{eq:HorneWarner},
such that the relation can retrieve $i$ using only $\Delta
\varphi_\text{orb}$ and the superhump-derived $q$. We use this relation
to constrain $i$ for HBHA 4204-09, SDSS-090113 and Gaia-593107, which
are eclipsing systems. We measured $\Delta\varphi_\text{orb}$ on
nSH-subtracted data.  

\mbox{HBHA 4204-09} has a deep eclipse, with \mbox{$\Delta
\varphi_\text{orb} = 0.049(4)$}. Using our value of \mbox{$q \approx
0.29(3)$}, we compute \mbox{$i = 77(1)\degr$}. On the other hand,
\mbox{SDSS-090113} shows grazing eclipses of variable depth and width
that disappear completely for some $\varphi_\text{prec}$. For this
reason, we can assume \mbox{$\Delta \varphi_\text{orb} \approx 0$}. With
our value of \mbox{$q \approx 0.47(4)$}, we compute \mbox{$i = 71\fdg
6(4)$}. For the last eclipsing system in our sample, \mbox{Gaia-593107},
we measure $\Delta \varphi_{\text{orb}} = 0.06(2)$. This eclipse width,
combined with $q \approx 0.38(8)$, gives an orbital inclination $i =
76(3)\degr$.
 
\section{Conclusions}\label{sec:Conclusions}

In this work we present results from a search for nSHs in poorly studied
CVs. We initially cross-matched TESS light-curve data with the VSX
catalogue for objects labelled as CV or NL. We manually inspected LS
periodograms of objects from this query ($n=180$), and then selected
targets with at least two neighbouring periodicities above the period
gap, including one large-period signal that matches the beating of the
former two. This resulted in six systems with nSHs, which we list in
Table~\ref{tab:shList}.

Spectroscopic measurements of $P_\text{orb}$ were available for only one
of the six aforementioned systems. For the rest, we used a couple of
methods to recognise $P_\text{orb}$ signatures in their LS periodograms.
Three systems had their orbital period determined by the presence of
eclipses. For the last two, $P_\text{orb}$ was identified using the
\mbox{$\varphi_\text{prec}$-dependent} irradiation of the secondary
caused by the precessing tilted disc (see Section
\ref{sec:superhump_degeneracy}). For all systems, the light maximum in
the orbital phase profile was found to shift to earlier
$\varphi_\text{orb}$, as $\varphi_\text{prec}$ advances. This is strong
evidence of nSHs \citep{Kimura2020, Kimura2021} and supports our
findings. For SDSS-090113, a peculiar behaviour was observed in ZTF
photometry (Figure~\ref{fig:star13special}), which is similar to what is
observed in IW~And stars \citep{Kato2019}. For Gaia-46843, two
Z~Cam-like episodes in ASAS-SN data were found, which take place after
drops in brightness of about 0.5 mag (Figure \ref{fig:star04special}).

Determined periodicities from TESS photometry can constrain some
physical parameters of CV systems. The dependence between the nSH
deficit $\varepsilon_-$ and the system mass ratio $q$ has been explored
in several works already (e.g. \citealp{Wood2009,Montgomery2009}).
Referenced $q(\varepsilon_-)$ relations, however, underestimate
independent measurements of $q$ and $\varepsilon_-$, which gives some
ground for concern (Figure~\ref{fig:qeps}(a)). We tried to come to a
better $q(\varepsilon_-)$ relation by using the precession rate of a
differentially rotating steady-state disc that extends to the maximum
tidal truncation radius $r_\text{tidal}$. This, in combination with the
$r_\text{tidal}(q)$ relation given in \citet[Table 1]{Paczynski1977},
resulted in better agreement with independent observations
(Figure~\ref{fig:qeps}(b)). Mass-ratio estimates using this method are
given in Table \ref{tab:shmassratios} for different disk radii
$R_d=0.9r_\text{tidal},1.0r_\text{tidal},1.1r_\text{tidal}$. For the
three systems that are eclipsing, we used the eclipse-mapping techniques
described by \citet{Horne1985,Warner1995} to compute the orbital
inclination $i$ with our values of $q$.

There are several subtle points that bear discussion. SAP light curves
from TESS-SPOC may contain instrumental effects that could affect LS
periodogram measurements; and photometry itself may be contaminated by
nearby sources. To address the former, we repeated our methods on PDCSAP
data, and found small differences in comparison to Table
\ref{tab:shList} measurements. Regarding the latter, we used
mean-subtracted fluxes in all analysis, which mitigates effects by
non-variable contaminating objects. None of our systems have bright
sources in their vicinity, except the case of Gaia-593107, which is
discussed in Section \ref{sec:gaia_crowded}.

Our variant of the nSH subtraction method in \citet{Kimura2021} shares
the same shortcomings. If the nSH profile is time-dependent, it cannot
be fully subtracted. In addition, the mass-transfer stream could happen
to obstruct some parts of the disc, causing the light maximum to occur
at earlier orbital phases $\varphi_\text{orb}$ \citep{Kimura2020}.
Inhomogeneities in the stellar surface brightness of the secondary can
produce similar effects, shifting the light maximum to earlier or to
later $\varphi_\text{orb}$.

The technique of using irradiation of the secondary in order to
determine $P_\text{orb}$ is entirely based on photometry, and
spectroscopic measurements of $P_\text{orb}$ could support its
feasibility. In addition, radial-velocity analysis would put constraints
on mass ratios, and would test the $q(\varepsilon_-)$ relation we
consider in Section \ref{sec:Discussion_massratio}. The newly discovered
systems in this work are therefore strongly encouraged for follow-up
spectroscopic observations.

\section*{Acknowledgements}

This work includes data collected by the TESS mission and made use of
\textsc{lightkurve}, a Python package for Kepler and TESS data analysis
\citep{lightkurve}. Funding for the TESS mission is provided by the
NASA's Science Mission Directorate. This work has made use of the
NASA/IPAC Infrared Science Archive, which is funded by the National
Aeronautics and Space Administration and operated by the California
Institute of Technology.

The CSS survey is funded by the National Aeronautics and Space
Administration under Grant No. NNG05GF22G issued through the Science
Mission Directorate Near-Earth Objects Observations Program. The CRTS
survey is supported by the U.S.~National Science Foundation under grants
AST-0909182 and AST-1313422.

We used the following Python packages for data analysis and
visualisation: \textsc{NumPy} \citep{numpy}, \textsc{SciPy}
\citep{scipy}, \textsc{pandas} \citep{pandas1,pandas2},
\textsc{Matplotlib} \citep{matplotlib} and \textsc{uncertainties}
\citep{uncertainties}. 

We are grateful to R.~K.~Zamanov and to A.~A.~Kurtenkov for their advice
during the preparation of this work. We thank the anonymous referee for
their time and their effort. We acknowledge the grants
\mbox{K$\Pi$-06-H28/2} and \mbox{K$\Pi$-06-M58/2} from the Bulgarian
National Science Fund. Both authors contributed equally to this work. It
is appreciated that our last names considerably simplified the issue of
author ordering.

\section*{Data Availability}
\label{sec:Data_Availability}

This work contains publicly available data from the sky surveys TESS,
ASAS-SN, CSS, CRTS, and ZTF, all of which can be found in their
corresponding databases.



\bibliographystyle{mnras}
\bibliography{bibliography} 




\appendix

\section{Extra material}
\begin{figure}
    \centering
    \includegraphics{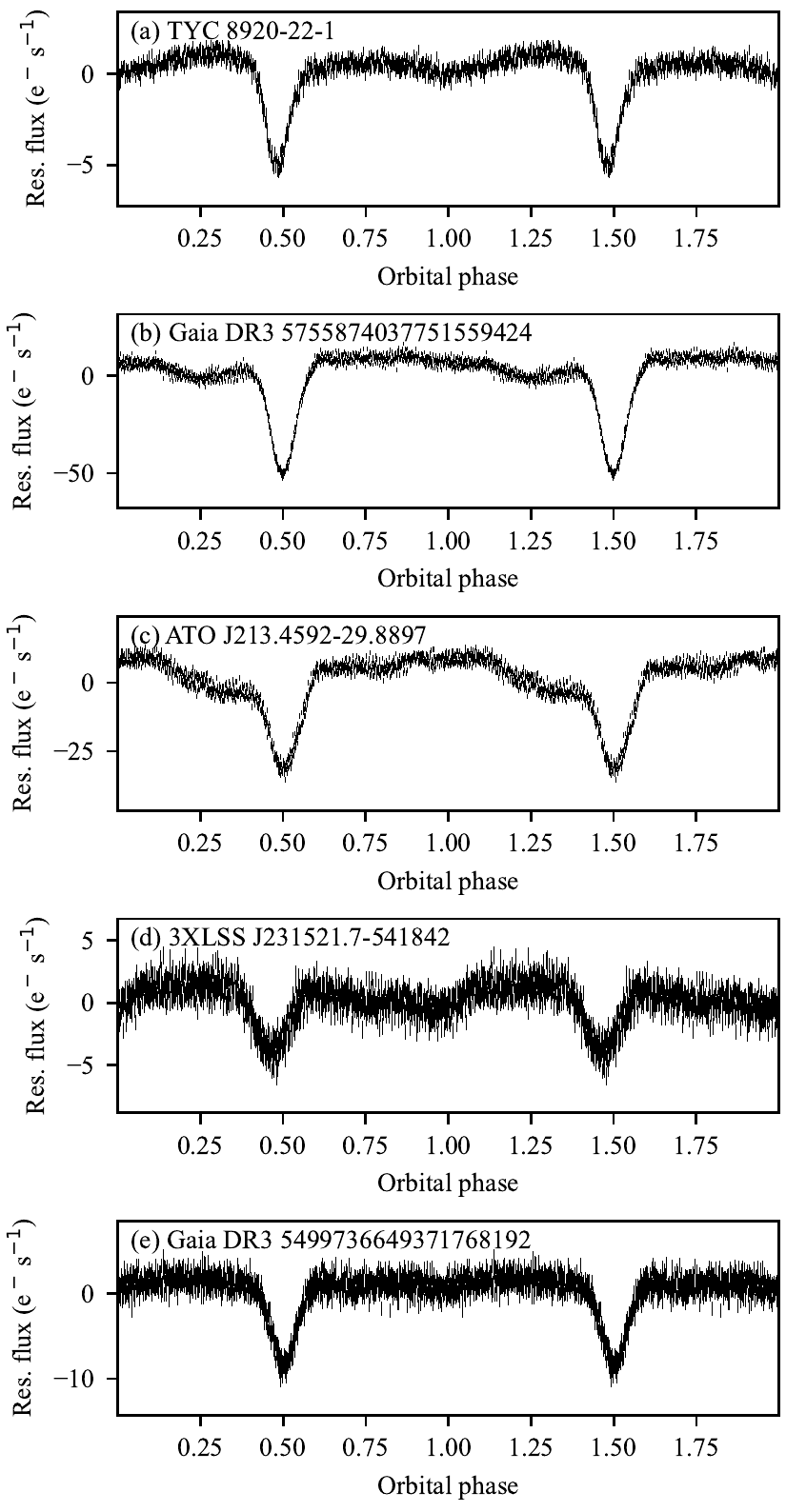}
    \caption{Orbital phase curves of Table~\ref{tab:ebList} systems.
    Light curves were smoothed by a fourth-order Savitzky-Golay filter
    of window size 10~d before folding. Orbital phases are offset with
    +0.50 for the sake of clarity.}
    \label{fig:ebsummary}
\end{figure}

This appendix contains orbital phase curves (Figure~\ref{fig:ebsummary})
and a list of $P_\text{orb}$ measurements (Table~\ref{tab:ebList}) of
discovered eclipsing binaries with no nSH behaviour. Our calculated
$q(\varepsilon_-)$ curves using the tidal truncation disc radii relation
$r_\text{tidal}(q)$ in \citet{Paczynski1977} are shown in
Table~\ref{tab:lookup_table}. Computed values can be approximated by
\begin{equation}\label{eq:e_q_appendix_fit}
    |\varepsilon_-| = \dfrac{a q}{(1+q)^{b}},
\end{equation}
where $a$ and $b$ are fit coefficients. Values of best fits are given in
Table~\ref{tab:lookup_table_fit} for
$R_d=0.9r_\text{tidal},1.0r_\text{tidal},1.1r_\text{tidal}$.

\begin{table}
\centering
\caption{Equation \eqref{eq:e_q_appendix_fit} fit coefficients $a$ and $b$ that approximate Table \ref{tab:lookup_table} data for different values of $R_d$. 
}
\label{tab:lookup_table_fit}
\begin{tabular}{cccc}
$R_d$ & $a$ & $b$ & $|\varepsilon_-|$ interval\\
\hline
$0.9 r_\text{tidal}$ & 0.155 & 1.706 & [0.006, 0.041] \\
$1.0 r_\text{tidal}$ & 0.182 & 1.694 & [0.006, 0.048] \\
$1.1 r_\text{tidal}$ & 0.210 & 1.680 & [0.006, 0.057]
\end{tabular}
\end{table}

\begin{table}
\centering
\caption{Tabular form of our derived $q(\varepsilon_-)$ relation for
$R_d = 0.9 - 1.1r_\text{tidal}$ using $r_\text{tidal}$ from \citet[Table
1]{Paczynski1977}.}
\label{tab:lookup_table}
\begin{tabular}{cccc}
\multirow{2}{*}{$|\varepsilon_-|$} & \multicolumn{3}{c}{$q(\varepsilon_-)$} \\[0.5ex]
\cline{2-4}\\[-1.5ex]
 & $0.9 r_\text{tidal}$ & $1.0 r_\text{tidal}$ & $1.1 r_\text{tidal}$
\\ \hline
0.006 & 0.039 & 0.032 & 0.031 \\
0.007 & 0.046 & 0.038 & 0.033 \\
0.008 & 0.053 & 0.044 & 0.038 \\
0.009 & 0.061 & 0.05  & 0.043 \\
0.010 & 0.069 & 0.057 & 0.048 \\
0.011 & 0.077 & 0.064 & 0.054 \\
0.012 & 0.086 & 0.071 & 0.060 \\
0.013 & 0.095 & 0.078 & 0.066 \\
0.014 & 0.104 & 0.085 & 0.071 \\
0.015 & 0.113 & 0.093 & 0.078 \\
0.016 & 0.124 & 0.101 & 0.084 \\
0.017 & 0.135 & 0.108 & 0.091 \\
0.018 & 0.145 & 0.117 & 0.097 \\
0.019 & 0.156 & 0.126 & 0.104 \\
0.020 & 0.167 & 0.135 & 0.110 \\
0.021 & 0.178 & 0.144 & 0.118 \\
0.022 & 0.191 & 0.153 & 0.126 \\
0.023 & 0.205 & 0.162 & 0.134 \\
0.024 & 0.218 & 0.171 & 0.142 \\
0.025 & 0.232 & 0.181 & 0.149 \\
0.026 & 0.246 & 0.193 & 0.157 \\
0.027 & 0.262 & 0.204 & 0.165 \\
0.028 & 0.279 & 0.216 & 0.173 \\
0.029 & 0.297 & 0.227 & 0.182 \\
0.030 & 0.314 & 0.239 & 0.192 \\
0.031 & 0.332 & 0.251 & 0.201 \\
0.032 & 0.352 & 0.265 & 0.211 \\
0.033 & 0.373 & 0.280 & 0.221 \\
0.034 & 0.393 & 0.295 & 0.231 \\
0.035 & 0.414 & 0.310 & 0.241 \\
0.036 & 0.436 & 0.324 & 0.251 \\
0.037 & 0.461 & 0.340 & 0.264 \\
0.038 & 0.487 & 0.358 & 0.277 \\
0.039 & 0.513 & 0.375 & 0.289 \\
0.040 & 0.539 & 0.392 & 0.302 \\
0.041 & 0.566 & 0.410 & 0.315 \\
0.042 &       & 0.427 & 0.328 \\
0.043 &       & 0.448 & 0.341 \\
0.044 &       & 0.470 & 0.356 \\
0.045 &       & 0.492 & 0.371 \\
0.046 &       & 0.514 & 0.386 \\
0.047 &       & 0.536 & 0.401 \\
0.048 &       & 0.558 & 0.415 \\
0.049 &       &       & 0.431 \\
0.050 &       &       & 0.449 \\
0.051 &       &       & 0.468 \\
0.052 &       &       & 0.487 \\
0.053 &       &       & 0.505 \\
0.054 &       &       & 0.524 \\
0.055 &       &       & 0.543 \\
0.056 &       &       & 0.562 \\
0.057 &       &       & 0.581
\end{tabular}
\end{table}

\begin{table*}
\caption{List of discovered eclipsing variables with no superhump
signatures. Equatorial coordinates come from Gaia DR3 and are in the
J2000 epoch.}
\label{tab:ebList}
\begin{tabular}{lrrcl}
\multicolumn{1}{c}{Name} &
\multicolumn{1}{c}{RA} &
\multicolumn{1}{c}{Dec} &
\multicolumn{1}{c}{TESS Sector} &
\multicolumn{1}{c}{P$_{\text{orb}}$}\\ \hline
TYC 8920-22-1 &
$ 06\h  56\m    39\fs    36$ &
$-67\degr 02' 16\farcs 8$ &
39 &
$0\fd 09081(18)$\\

Gaia DR3 5755874037751559424 &
$ 09\h  02\m    57\fs    71$ &
$-07\degr 59' 19\farcs 9$ &
$35$ &
$0\fd 15585(53)$\\

ATO J213.4592-29.8897 &
$ 14\h  13\m    50\fs    22$ &
$-29\degr 53' 23\farcs 1$ &
$38$ &
$0\fd 13879(43)$\\

3XLSS J231521.7-541842 &
$ 23\h  15\m    21\fs    72$ &
$-54\degr 18' 43\farcs 1$ &
$28$ &
$0\fd 14966(54) $\\

Gaia DR3 5499736649371768192 &
$ 06\h  27\m    51\fs    08$ &
$-53\degr 45' 17\farcs 7$ &
39 &
$0\fd 15838(53)$\\
\end{tabular}
\end{table*}

\bsp	
\label{lastpage}
\end{document}